# The Large-Number Coincidence, the Cosmic Coincidence and the Critical Acceleration

*Scott Funkhouser*
Joint Sciences Dept. of the Claremont Colleges, 925 N. Mills Rd. Claremont, CA 91711[*]
[*]Current address of affiliation: Dept. of Physics, The Citadel, 202 Grimsley Hall, Charleston, SC 29409

ABSTRACT
The coincidence problem among the pure numbers of order near $10^{40}$ is resolved with the Raychaudhuri and Friedmann-Robertson-Lemaitre-Walker equations and a trivial relationship involving the fine structure constant. The fact that the large number coincidence occurs only in the same epoch in which other coincidences among cosmic parameters occur could be considered a distinct coincidence problem suggesting an underlying physical connection. A natural set of scaling laws for the cosmological constant and the critical acceleration are identified that would resolve the coincidence among cosmic coincidences.

## 1. Introduction

Eddington [1] and Dirac [2] hypothesized that some unidentified physics was responsible for the ensemble of pure numbers of order near $10^{40}$ that could be generated from the parameters of the universe. While a host of explanations have been offered the problem of the Large-Number Coincidence (LNC) has persisted over the decades. Dirac suggested that the LNC would be resolved if the Newtonian gravitation constant were to vary in time [3]. Modern discussions of the coincidence have included considerations of anthropic selection mechanisms [4] and the holographic *N*-bound conjecture [5]. However, the coincidence problems among the parameters of nature have only multiplied since the time of Dirac and Eddington with the discovery of other, apparently separate coincidences among cosmological parameters that seem to occur only this epoch.

In Section 2 of this letter the LNC is resolved using physical scaling laws from the standard cosmological model that were not known in the early 20th century. However, there is a lingering problem associated with the LNC in that it occurs only in this epoch in which other coincidences involving cosmological parameters occur. In Section 3 this new problem is discussed and natural scaling laws for the cosmological constant and the critical acceleration are identified that would resolve the problem. In the remainder of this section the terms constituting the large number coincidence are presented.

Many of the large pure numbers of order near $10^{40}$ involve cosmological parameters such as the mass, radius and age of the observable universe. According to the interpretation of data from the recent WMAP observations, the Big Bang occurred around $T_0 \approx 4.3 \times 10^{17}$ seconds ago [6]. (Only two significant figures will be employed in this letter.) The conformal time $T_0$ corresponds to a cosmic particle horizon $R_0 \approx 1.3 \times 10^{26}$m. This distance $R_0$ is the term traditionally used to represent the size of the observable universe [7]. The current average density $\varepsilon_0$ of energy in the observable universe is very close to the critical density $c^2\rho_c = c^2 \cdot 3H_0^2(8\pi G)^{-1} \cong c^2 \cdot 9.6 \times 10^{-27}$kg m$^{-3}$, where $H_0 \approx 2.3 \times 10^{-18}$s$^{-1}$ is the current value of the Hubble parameter [6]. Matter comprises approximately $\Omega_m \approx 27$ percent of the total density and baryonic matter comprises just $\Omega_b \approx 4.4$ percent of the total [6]. Some form vacuum energy, perhaps due to a cosmological constant $\Lambda$, comprises $\Omega_\Lambda \approx 73$ percent of the total cosmic energy density [6]. Given that spacetime is flat on the largest scales the total mass $M_0$ of the observable universe can be calculated as $M_0 = \Omega_m \cdot (4\pi/3)R_0^3 \cdot \rho_c$ which gives $M_0 \cong 2.4 \times 10^{52}$ kg.

In the time of Dirac and Eddington at least four pure numbers of order near $10^{40}$ were associated with the LNC:

$$R_0/r_e \approx 4.6 \times 10^{40}, \tag{1}$$

$$cT_0/r_e \approx 4.6 \times 10^{40}, \tag{2}$$

$$e^2/(Gm_n m_e) \approx 2.3 \times 10^{39}, \tag{3}$$

$$(M_0/m_n)^{1/2} \approx 3.8 \times 10^{39}, \tag{4}$$

where $c$ is the speed of light in a vacuum, $e$ is the unit of fundamental electric charge, $G$ is the constant of gravitation, $m_n$ and $m_e$ are the masses of the nucleon and electron, respectively, and $r_e$ is the classical radius of the electron. Note that since baryonic matter comprises just $\Omega_b/\Omega_m \approx 16$ percent of all matter Eq. (4) is not exactly the square root of the baryon number of the observable universe. Since the time of the early investigations at least three other pure numbers of order near $10^{40}$ have entered the discourse [9]:

$$(m_P/m_n)^2 \approx 1.7 \times 10^{38}, \tag{5}$$

$$(M_0/m_P)^{2/3} \approx 1.1 \times 10^{40}, \tag{6}$$

$$R_0/\lambda_n \approx 9.8 \times 10^{40}, \tag{7}$$

where $\lambda_n \equiv h/(m_n c)$ is the Compton wavelength of the nucleon, $m_P \equiv \sqrt{\hbar c/G}$ is the Planck mass and $\hbar \equiv h/(2\pi)$ is the reduced Planck constant. The terms in (1) – (7) will be considered to constitute the large number coincidence.

## 2. Resolving the Large Number Coincidence

Three seemingly external equations are found to reduce the gallery of large pure numbers in (1) – (7) to just three that do not constitute a coincidence problem. Two of the reducing equations follow from the Raychaudhuri equation. The Raychaudhuri equation is an important formulation of the motion of cosmological matter that is consistent with the general theory of relativity but may be derived independently [10]. During the era of matter-dominance the Raychaudhuri equation leads to the scaling law

$$H \sim c/R \sim 1/T, \tag{8}$$

where $H$ is the Hubble parameter and $T$ is the conformal time associated with the co-moving particle horizon $R$. Eq. (8) is still roughly satisfied in this epoch since the era of matter-dominance ended only recently (cosmologically speaking) and it is one of the equations employed to resolve the large number coincidence.

A second resolving equation is obtained from Eq. (8) and some basic relationships from the standard cosmological model. According to the Friedmann-Robertson-Lemaitre-Walker equations, the Hubble parameter in a universe with zero curvature is related to the average total energy density $\varepsilon$ by [11]

$$H^2 = \frac{8\pi G}{3c^2}\varepsilon. \tag{9}$$

During the era of matter-dominance the total energy density is $\varepsilon \approx c^2 \rho_m$ where $\rho_m$ is the density of matter. Using Eq. (9) to express the density of matter in terms of the Hubble parameter, the total mass $M$ contained within the observable Universe would be

$M=R^3H^2/(2G)$. With a substitution from Eq. (8), the following scaling law is obtained that is also the second relationship needed to resolve the LNC:

$$\frac{GM}{R} \sim c^2. \tag{10}$$

Like Eq. (8), Eq. (10) is still roughly satisfied in this epoch even though the energy density of the universe is no longer matter-dominated. The left side of Eq. (10) is approximately equal to $1.2 \times 10^{16} m^2 s^{-2}$ at this time and its proximity to $c^2$ was regarded as a suggestive coincidence among large numbers before it was known to be a scaling law and before detailed knowledge of cosmological parameters was available [8], [12].

The third equation employed to resolve the LNC is most likely just a simple coincidence:

$$(m_e/m_n) \approx 5.5 \times 10^{-4} \sim 10^{-1}\alpha, \tag{11}$$

where $\alpha \approx 7.3 \times 10^{-3}$ is the fine structure constant. In the context of large pure numbers of order near $10^{40}$ the factor of ten may be ignored and it can be said that the fine structure constant is roughly of order near the ratio of the electron mass to the nucleon mass.

Using Eqs. (8), (10) and (11) the gallery of large pure numbers in (1) – (7) may be reduced as follows. First, Eq. (11) causes the Compton length of the nucleon to have the same order of magnitude as the classical radius of the electron, and the term in (7) thus follows from (1). Eq. (2) also follows immediately from (1) due to Eq. (8). Eq. (3) follows from Eqs. (1), (4) and (10). Finally, Eq. (1) follows from Eqs. (4), (5), (10) and (11). The only remaining irreducible large pure numbers of order near $10^{40}$ are therefore those in Eqs. (4), (5) and (6):

$$\left(\frac{M_0}{m_n}\right)^{1/2} \sim \left(\frac{m_P}{m_n}\right)^2 \sim \left(\frac{M_0}{m_P}\right)^{2/3}. \tag{12}$$

The group of terms in (12) does not constitute a coincidence problem, and the LNC can be said to have been resolved with Eqs. (8), (10) and (11). Dirac and Eddington were therefore correct to hypothesize that some implicit physics was involved in generating the coincidence problem. Note that the terms in (12) may be connected by some yet-unspecified physics. For instance, the Eddington-Weinberg relation

$$H_0 h^2 \sim Gcm_n^3 \tag{13}$$

follows from Eqs. (8), (10) and (12), and the holographic $N$-bound conjecture may provide some insight toward explaining that relationship [5]. Also, anthropic selection mechanisms could perhaps be identified that would influence the relationships in Eq. (12) or others among cosmological parameters.

## 3. The Coincidence of Cosmic Coincidences

It is important to note that, while the LNC is resolved with Eqs. (8), (10) and (11), the numerical coincidence problem associated with it could occur only in this cosmological epoch. That fact may create a distinct coincidence problem when considered in conjunction with other coincidence problems that are unique to this time. It is considered remarkable that, only in this epoch, the energy density of matter in the universe is of order the vacuum energy density attributed to the cosmological constant $\Lambda$. This coincidence is known as the Cosmic Coincidence and may be expressed as

$$\Omega_m \rho_c \sim \frac{3\Lambda}{8\pi G}, \tag{14}$$

where the term on the left is approximately $2.3 \times 10^{-10}$ J m$^{-3}$. The cosmological constant is approximately $\Lambda \approx 3.9 \times 10^{-36}$ s$^{-2}$ and corresponds to a vacuum energy density $\varepsilon_{vac} = 3\Lambda c^2/(8\pi G) \cong 6.2 \times 10^{-10}$ J m$^{-3}$. A fundamental scaling that is responsible for the value of $\Lambda$ has not been derived.

So, this cosmological epoch could be special for at least two presumably distinct reasons: the LNC and the Cosmic Coincidence occur only in this epoch. It may be more reasonable to hypothesize that an underlying physical connection is implicit in the simultaneous occurrence of two, presumably distinct coincidence problems among similar terms rather than to stipulate that only chance is responsible. A scaling law that could explain the coincidence of cosmic coincidences, as it were, is:

$$\Lambda \sim \left(\frac{8\pi G}{3c^2}\right) \frac{Gm_n^2}{\lambda_n^4}. \tag{15}$$

That is to say that the energy density associated with the cosmological constant may be scaled to the gravitational energy density of the nucleon mass confined to a sphere whose radius is the Compton wavelength of the nucleon. The right side of Eq. (15) is approximately equal to $3.7 \times 10^{-31}$ s$^{-2}$, but replacing $\lambda_n$ with $b\lambda_n$ where $b \sim 10$ could reasonably account for the difference between the right and left sides. The relationships in Eq. (12) would follow from Eq. (15) and the Cosmic Coincidence, and the LNC would thus naturally result from the Cosmic Coincidence. Also note that Eq. (15) follows immediately from (13) and the fact that $H_0 \sim \Lambda^{1/2}$, which is a result of Eq. (9) and vacuum-dominance.

There may be yet another curious coincidence among cosmological parameters that is unique to this epoch. The observed motions of clusters of galaxies and material within galaxies may be interpreted to indicate that the laws of dynamics deviate from Newtonian models at accelerations smaller than some critical acceleration $a_0 \sim 10^{-10}$ m s$^{-2}$ [13]. It so happens that the Hubble acceleration $H_0 c$ is of order $10^{-10}$ m s$^{-2}$ only in this epoch. With the substitution $H_0 \sim \Lambda^{1/2}$ the coincidence of the critical acceleration may be expressed in the more suggestive form

$$a_0 \sim c\sqrt{\Lambda}. \tag{16}$$

The coincidence $a_0 \sim H_0 c$ is well known [13]. The point of this present analysis is that, insofar as it represents a problem, the coincidence of the critical acceleration generates a distinct coincidence problem in that it occurs only in the same epoch in which the Cosmic Coincidence and the LNC occur. However, if the critical acceleration were scaled to the characteristic gravitational acceleration of the nucleon mass at its Compton length

$$a_0 \sim \frac{Gm_n}{\lambda_n^2}, \tag{17}$$

which follows from Eqs. (15) and (16), then the coincidence of the critical acceleration would follow from the Cosmic Coincidence. The right side of Eq. (17) is approximately $6.3 \times 10^{-8}$ m s$^{-2}$, but as with Eq. (15), replacing $\lambda_n$ with $b\lambda_n$ where $b \sim 10$ could reasonably account for the difference between the right and left sides. The proposed scaling laws in Eqs. (15) and (17) may also be linked to the holographic $N$-bound conjecture [5].

**Acknowledgments**: This work benefited from discussions with Lloyd Knox, Moti Milgrom, Tom Mongan and Joe Tenn.